\documentstyle[psfig,referee]{mn}

\newcommand{\ha}{\ifmmode {\rm H}\alpha \else H$\alpha$\fi}
\newcommand{\hb}{\ifmmode {\rm H}\beta \else H$\beta$\fi}
\newcommand{\lya}{\ifmmode {\rm Ly}\alpha \else Ly$\alpha$\fi}
\newcommand{\lyb}{\ifmmode {\rm Ly}\beta \else Ly$\beta$\fi}
\newcommand{\ciii}{\ifmmode {\rm C}\,{\sc iii} \else C\,{\sc iii}\fi}
\newcommand{\civ}{\ifmmode {\rm C}\,{\sc iv} \else C\,{\sc iv}\fi}

\newcommand{\kms}{\rm\ km\ s^{-1}}

\def\kms{\ifmmode {\rm \ km \ s^{-1}}\else \rm km \ s^{-1}\fi}
\def\cm{\ifmmode {\rm \ cm }\else $\rm cm$\fi}
\def\Mpc{\ifmmode {\rm \ Mpc }\else $\rm Mpc$\fi}
\def\kpc{\ifmmode {\rm \ kpc }\else $\rm kpc$\fi}
\def\yrs{\ifmmode {\rm \ yrs }\else $\rm yrs$\fi}
\def\erg{\ifmmode {\rm \ erg }\else $\rm erg$\fi}
\def\sec{\ifmmode {\rm \ s }\else $\rm s$\fi}
\def\ergs{\ifmmode {\rm \ ergs }\else $\rm ergs$\fi}

\title{P-Cygni Type Ly$\alpha$ from Starburst Galaxies}

\author[Ahn, Lee, \& Lee]
  {Sang-Hyeon Ahn,$^1$ Hee-Won Lee$^2$ and Hyung Mok Lee$^3$\\
$^1$ School of Physics, Korea Institute for Advanced Study, 207-43 Cheongyangri-dong, Dongdaemun-gu, Seoul, 130-012, Korea \\
$^2$ Department of Geoinformation Sciences, Sejong University, Seoul 143-747, Korea\\
$^3$ School of Earth and Environmental Sciences, Astronomy Program, 
Seoul National University, Seoul 151-742, Korea.}

\date{Submitted version, 29 Mar 2002}

\begin{document}

\maketitle

\begin{abstract}
P-Cygni type Ly$\alpha$ profiles exhibited in nearly half of 
starburst galaxies, both nearby and high-z, are believed to be formed 
by an expanding supershell surrounding a star-forming region.
We apply the Monte Carlo code which was developed previously
for static and plane-parallel medium
to calculate the Ly$\alpha$ line transfer in a supershell of 
neutral hydrogen which are expanding radially in a spherical bulk flow.
We consider typical cases that the supershell has the Ly$\alpha$ 
line-centre optical depth of $\tau_0=10^5-10^7$, a radial expansion velocity 
of $V_{\rm exp} \sim 300 \kms$, and the turbulence of $b \simeq 40 \kms$.
We find that there appear a few emission peaks at the
frequencies corresponding to $(2N-1)V_{\rm exp}$, 
where the order of back scatterings $N > 1$.
As $V_{\rm exp} \rightarrow b$, the emergent profiles become
similar to those for the static medium and the peaks are less prominent.
We also investigate the effects of column density 
of the supershell on the emergent Ly$\alpha$ profiles. 
We find that the number and the flux ratios of emission peaks
are determined by interplay of $\tau_0$ and $V_{\rm exp}$
of the supershell. We discuss the effects of dust extinction 
and the implication of our works 
in relation to recent spectroscopic observations of 
starburst galaxies.
\end{abstract}


\section{Introduction}

Ly$\alpha$ emission lines render us a very useful redshift 
indicator for starburst galaxies at high redshift, because
their rest-frame equivalent width was predicted to be 
$50-200{\rm\ \AA}$ (Charlot \& Fall 1993).
Thommes et al. (1998) suggested that the existence of strong Ly$\alpha$
emission is a necessary condition for primeval galaxies.
However, dust enrichment is believed to be a fast process
in the early evolution of galaxies, and therefore Ly$\alpha$ 
should suffer severe extinction by dust.
It has been argued that even a small amount of dust in a very 
thick hydrogen medium can destroy Ly$\alpha$ photons because 
their path lengths before escape increase tremendously 
due to a huge number of resonance scatterings.
Hence, dust absorption is believed to be the main reason 
for the early null detection of Ly$\alpha$ emission from 
high-$z$ galaxies by narrow band imaging surveys.

The Lyman break method has been very successful among a few programs
to detect high-$z$ galaxies. Steidel et al. (1996) presented the 
rest-frame UV spectra of primeval galaxies at 
$\langle z \rangle=3$. They continued to
search for even more distant galaxies, and secured the spectra of
galaxies at the redshift $\langle z \rangle=4$
(Steidel et al. 1999).
The UV spectra of those primeval galaxies are characterized by a flat
UV continuum and the three types of the Ly$\alpha$ feature: 
symmetric emission, broad absorption in the wings, and P-Cygni type
asymmetric emission.

From {\it HST GHRS} observations
of nearby H II galaxies, Kunth et al. (1998) discovered
that P-Cygni type Ly$\alpha$ emission lines are always accompanied by
the low-ionization interstellar absorption lines blueshifted 
by $100-300 \kms$ with respect to the systemic velocity,
while broad wing absorptions of Ly$\alpha$ are accompanied by
the much less blueshifted interstellar absorption lines.
There are spectral similarities between 
nearby and high-$z$ starburst galaxies.
Recently there have been a number of observations to
obtain rest-frame optical nebular lines from high-$z$ galaxies
(Teplitz et al. 2000a; Teplitz et al. 2000b; Pettini et al. 2001).

These observations support the idea 
that the kinematics of surrounding media plays a crucial role in
the survival of Ly$\alpha$ photons in primeval galaxies,
which is similar to the case in nearby starburst galaxies.
When the systemic redshifts were determined by using the nebular lines,
evidence for bulk motions of several hundred $\kms$ is found 
(Pettini et al. 2001). Ahn \& Lee (1998) and Lee \& Ahn (1998) 
investigated the formation of asymmetric Ly$\alpha$ line profiles 
observed in the spectra of high-z galaxies 
by introducing the galactic superwind model.

It is generally believed that the Ly$\alpha$
emission is not a good indicator for the star-formation rate,
because of dust extinction.
However, Ly$\alpha$ emission often bears 
a nearly unique information on high-z starburst galaxies.
Therefore, we have to understand the line formation mechanism,
in order to establish the quantitative measure of star-formation rate.

We have been investigating the Ly$\alpha$ line formation in thick
and static media of neutral hydrogen, and developed an accurate and efficient
Monte Carlo method (Ahn, Lee, \& Lee 2000, 2001, 2002). In this paper
we apply our method to an expanding spherical supershell.
In section 2 we describe our Monte Carlo code. 
We show the detailed mechanism of Ly$\alpha$ line transfer
in expanding media in section 3, and the effects of both thickness 
and expansion velocity of the supershell in section 4. 
In the last section we summarize the 
results and discuss on the implication of our works.

\section{Monte Carlo Method}

\begin{figure*}
\begin{center}
\parbox{1cm}{
\psfig{file=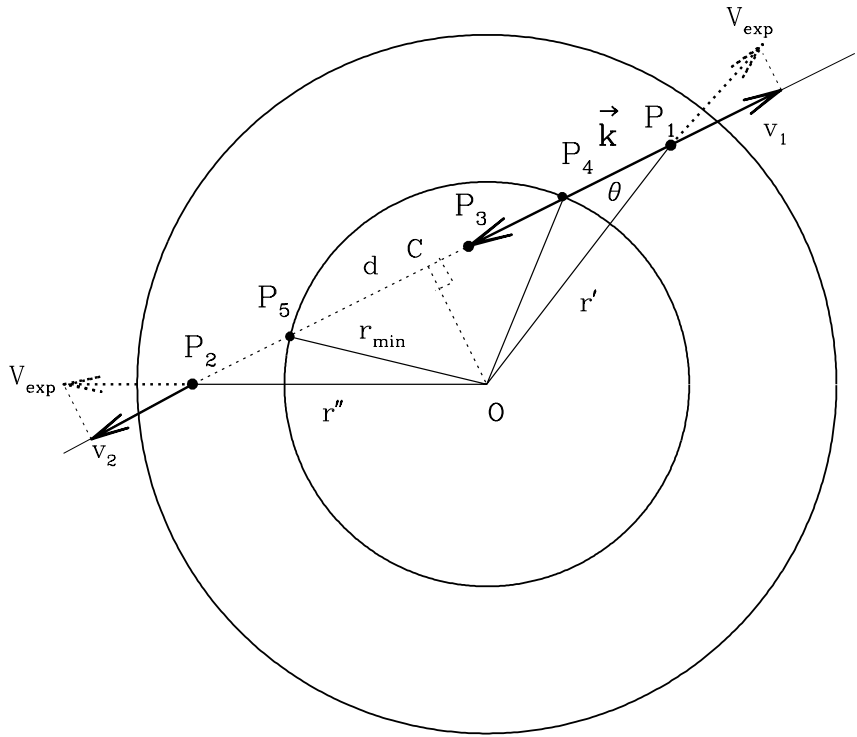,height=14cm}}
\caption{Schematic diagram for the back-scattered photons.
Here a Ly$\alpha$ photon would propagate from $P_1$ to $P_3$.
However, because the central region is devoid of neutral hydrogen,
the photon traverse freely to $P_2$, which is receding with respect to
the photon. Details are explained in the text.}
\end{center}
\end{figure*}

Our model for starburst galaxies is illustrated 
in Figure~1. We assume that there is a central star-forming
region surrounded by an expanding spherical supershell.
We also assume that the Ly$\alpha$ source is the central super 
star cluster, because the Ly$\alpha$ luminosity of a typical 
starburst galaxy is equivalent to ionizing photons 
from $10^3-10^4$ O, B stars (Gonz\'alez Delgado 1998; Ahn 2000)
and the star-forming region is compact (Frye et al. 2002).

Expanding supershells are frequently seen in nearby starburst
galaxies in a kiloparsec scale around the central star-forming
region (Marlowe et al. 1995; Martin 1998).
Recent {\it HST} observations revealed that the neutral column
density of a supershell ranges $N_{\rm HI}\sim10^{18}-10^{22}\cm^{-2}$,
which is deduced from usual Voigt fitting analyses (Kunth et al. 1998).
The Ly$\alpha$ line-centre optical depth is related with the H I column 
density $N_{\rm HI}$ by 
\begin{equation}
\tau_0 \equiv 1.41\ \left({b \over 12.9 \kms}\right)^{-1}
\left[{N_{\rm HI} \over {10^{13} \rm cm^{-2}}}\right],
\end{equation}
where $b$ is the Doppler width for turbulence motion.
Therefore, we should deal with scattering media with the line-centre
optical depths $\tau_0=10^5-10^9$.
However, Monte Carlo methods are usually inefficient 
in this regime, because of the large computing time 
due to the large number of resonance scatterings required before escape.
Hence we developed an accelerated Monte Carlo 
method in the previous paper (Ahn, Lee, \& Lee 2002).

In an expanding supershell, some photons are scattered
back into the opposite part of the supershell, and so they get redshifted.
They experience a series of such back-scatterings 
before they escape the system. 
These processes enhance the escape probability of Ly$\alpha$ 
resonance photons in optically thick media than in static medium,
and the number of scatterings before escape becomes much less than 
that in static media.
Therefore, even though we want to solve the problem of radiative
transfer in extremely thick media, it is sufficient for us to
use the original code described in Ahn et al. (2001).

We illustrate the back-scattering process in Figure~1. Suppose that
a photon scattered at $P_1$ 
to a direction ${\bf k} \parallel \overrightarrow{P_1P_3}$
is scattered at a next scattering site $P_3$ if 
the central region is also uniformly filled with neutral hydrogen
atoms. Since no neutral hydrogen is assumed to be present 
in $\overline{P_4P_5}$, the photon traverses to $P_2$, 
where $|\overline{P_5P_2}|=|\overline{P_4P_3}|$.
We determine the distance $2d\equiv |\overline{P_4P_5}|$ along the wave
vector ${\bf k}$, and add $|\overline{P_4P_5}|$ to the path length 
$|\overline{P_1P_3}|$ in the code. Here we define the position vectors 
of the adjacent scatterings to be $\bf{r^\prime}$ and $\bf{r^{\prime\prime}}$, 
$r^\prime = |{\bf r^{\prime}}|$,
and the angle $\theta \equiv \angle{OP_1P_4}$.
An elementary geometrical calculation gives
\begin{equation}
d = r^\prime\ \cos\theta - l,
\end{equation}
where
\begin{equation}
\cos\theta= -{{\bf r^\prime} \cdot {\bf k} \over |{\bf r}^\prime|},
\end{equation}
$l\equiv |\overline{P_1P_4}| = r^\prime\cos\theta-
\left[r^{\prime2}\cos^2\theta-r^{\prime2}+r_{min}^2\right]^{1/2}$,
and $|\overline{OP_4}|=|\overline{OP_5}|\equiv r_{min}$.

The velocity shifts due to the bulk expansion of the
medium should be taken into account. Referring to the geometry shown
in Figure~1, the velocities ${\bf v_1}$ and ${\bf v_2}$ are given by
\begin{equation}
{\bf v_1} = - V_{\rm exp} \cos\theta\ {\bf k},
\end{equation}
and
\begin{equation}
{\bf v_2} = V_{\rm exp} {s - l + d \over r^{''}}\ {\bf k},
\end{equation}
where $V_{\rm exp}$ is the expansion velocity of the supershell
and $V_{\rm exp} > 0$ by definition.
Here $s\equiv |\overline{P_1P_3}|$ is the path length 
corresponding to an optical depth $\tau=-\ln R$, where $R \in[0,1]$
is a random number.
We subtract frequency corresponding to the relative 
velocity ${\bf \Delta v= v_2-v_1}$
from the frequency $x$ whenever a back-scattering occurs.
Here we introduce the dimensionless frequency parameter $x$
defined by
\begin{equation}
x \equiv {\Delta\nu \over \Delta\nu_D}
={(\nu-\nu_0)\over \Delta\nu_D},
\end{equation}
which describes the frequency shift from the line-centre frequency $\nu_0$
in units of the Doppler width $\Delta\nu_D\equiv \nu_0 (b/c)$.
Here $b$ is the Doppler parameter that combines both turbulence and
the thermal motions of the scattering medium, and
$c$ is the speed of light.
The detailed description of our code can be found in the 
previous papers (Ahn, Lee, \& Lee 2000, 2001, 2002).

While scatterings occur in the supershell, 
we should also take into account the Doppler shift caused by
the relative bulk motion between the two adjacent scattering positions.
For the radial vectors of the adjacent scattering locations, $\bf r_1$
and $\bf r_2$, the expansion velocities projected
into the wave vector should be considered.
After a simple geometrical consideration, we obtain
\begin{equation}
\Delta v = V_{\rm exp}
\left({{\bf r_2} \over |{\bf r_2}|}
  -   {{\bf r_1} \over |{\bf r_1}|}\right) \cdot {\bf k}.
\end{equation}
We subtract this shift from the frequency of a Ly$\alpha$
photon whenever scatterings take place in the medium.
When a Ly$\alpha$ photon escapes the medium by a back-scattering,
we consider the redshifts due to the final back-scattering;
i.e., only $\bf v_1$ in Eq.(4) is taken into account.
We also calculate the path lengths of photons traversing in the
supershell in order to estimate the amount of dust extinction.
However, dust effect is considered very briefly in this paper,
and we will concentrate on the Ly$\alpha$ line transfer in the 
expanding supershell.

\section{The Formation of P-Cygni Type Ly$\alpha$ Profiles}

%%%%%%%% FIG 2
\begin{figure*}
\begin{center}
\parbox{1cm}{
\psfig{file=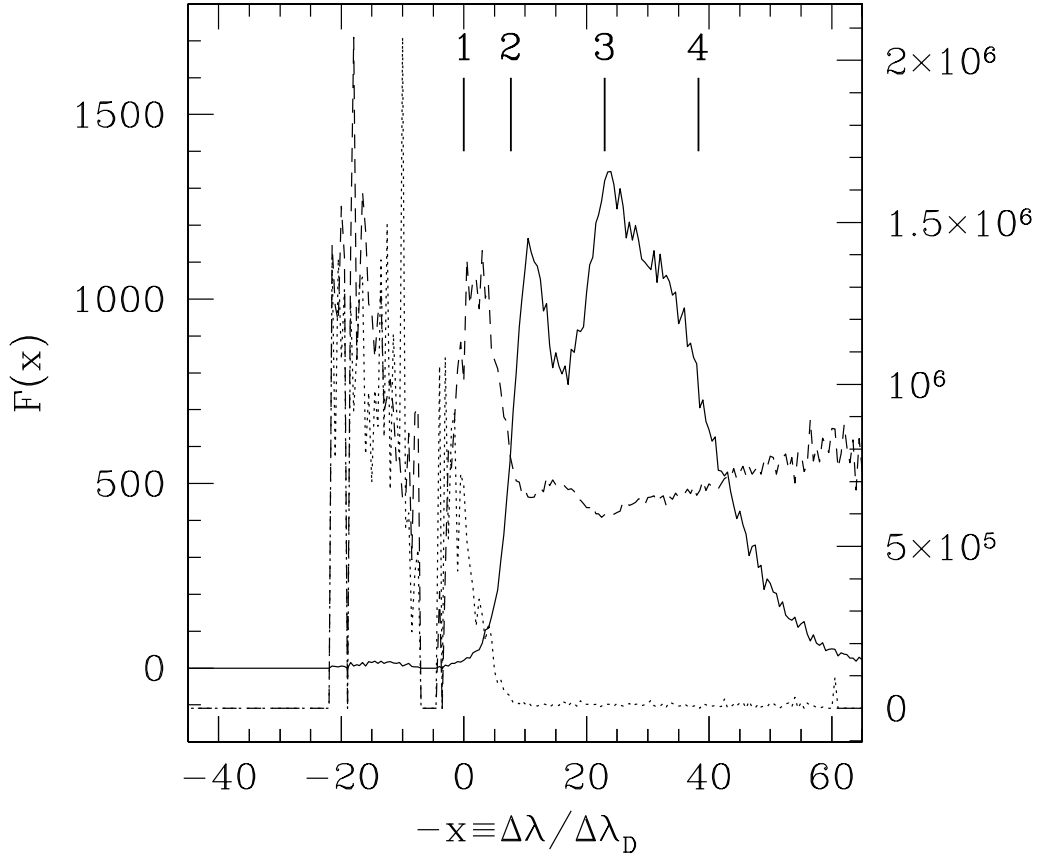,height=14cm}}
\caption
{An emergent profile (solid line) of Ly$\alpha$ emission calculated 
by our Monte Carlo code. The dashed line represents the total number
of scatterings before escape, and the dotted line refers to the
number of successive wing scatterings just before escape.
The horizontal axis refers to the wavelength shift in units of
thermal width. The right vertical axis refers to the number of 
scatterings, and the left one refers to the number flux of
the emergent spectrum. The vertical lines with numbers in the box
indicate the frequencies of possible emission peaks.
See the model parameters in the text.}
\end{center}
\end{figure*}
%%%%%%%%%%%%%%%%%%%%%

We now discuss on typical profiles of Ly$\alpha$
resulting from the scattering processes discussed above.
The expanding supershells in starburst galaxies have typical turbulence 
which is given by the Doppler parameter, $b \simeq 40 \kms$ estimated 
from Voigt fittings ({\it e.g.} Kunth et al. 1998). 
Hence we assume that the Voigt parameter of the expanding 
supershell to be $a=1.5\times 10^{-4} (40 \kms/b)$,
where the Voigt parameter $a \equiv \Gamma / 4\pi \Delta \nu_D$
with $\Gamma$ being damping constant. Here we neglect the thermal motion, 
because the turbulent motion usually exceeds the thermal motion. 
The FWHM of the input profiles is set to 
be $\Delta x = 2$, because the optical nebular emission 
lines indicate that $\Delta v = 70 \sim 100 \kms$ (Pettini et al. 2001).
We consider a supershell with its expansion velocity 
being $V_{\rm exp}\sim 300 \kms$
and its column density $10^{18} \cm^{-2} < N_{\rm HI} < 10^{21} \cm^{-2}$,
in accordance with observations in nearby starbursts (Kunth et al. 1998)
and high-z galaxies (Pettini et al. 2001).
These two quantities satisfy the condition $a\tilde\tau_0>10^3$, 
which guarantees the validity of the analytic solution
given by Adams (1972), Harrington (1973), and Neufeld (1990).
Here we define the line centre-optical depth by 
$\tilde\tau_0 \equiv \sqrt{\pi} \tau_0$, which
is a little bit different from the previous researches.

In Figure~2 we show the emergent profile for the model.
The number of photons used in the Monte Carlo calculation
is 80,000, and the other physical parameters are assumed 
to be the typical values mentioned above. 
Especially we choose the neutral column 
density $N_{\rm HI}=2\times10^{20}\rm cm^{-2}$.
The dotted line in the figure represents the average 
number of scatterings for a photon during its transfer, 
and the dashed line represents the average number of successive 
wing scatterings just before its escape. 
The solid line refers to the emergent Ly$\alpha$ profile.
The vertical bars with numbers refer to the frequency
of emission peaks corresponding to $N$-th back-scattering.
We can see two kinds of peaks in the profile:
one located at the bluer frequency, and the other three 
at the redder part. 

We see in the figure that the weak peak at the blue part is composed of
photons which have experienced very large number of scatterings
in the wings as well as in the core.
According to Adams (1972), the number of scatterings
for those photons is $N \propto \tau_0$.
Hence we can see that these are photons transferred in the supershell 
by so-called excursions (Adams 1972).
Some photons around $x \simeq 0$ are also similar ones, 
which is evident by the large number of scatterings
and wing scatterings. 

These weak peaks are formed through the following processes.
When photons with the line-centre frequency are incident upon
the inner wall of the expanding supershell, its scattering is 
off-centred by an amount of $x_{\rm exp}$, where we define the dimensionless
frequency corresponding to the expansion velocity 
$x_{\rm exp} \equiv - V_{\rm exp} / v_{\rm th}$.
Hence, the fraction of transmitted photons through
the supershell is approximately given by Eq.(2.23) in Neufeld (1990),
and the profile of transmitted photons is given by Eq. (2.32) in 
the same paper. Since $x_{\rm exp}=-7.65$ in the above simulation, 
we can see from Eq.(2.23) in Neufeld (1990) that a very small amount 
of photons can transmit the optically very thick supershell and 
majority of photons are back-scattered. The transmitted photons 
are redistributed into both sides of $x=-x_{\rm exp}$.
The radiative transfer of those photons was studied
very well both analytically (Neufeld 1990) and numerically
(Ahn et al. 2002).

%%%%%%%% FIG 3
\begin{figure*}
\begin{center}
\parbox{1cm}{
\psfig{file=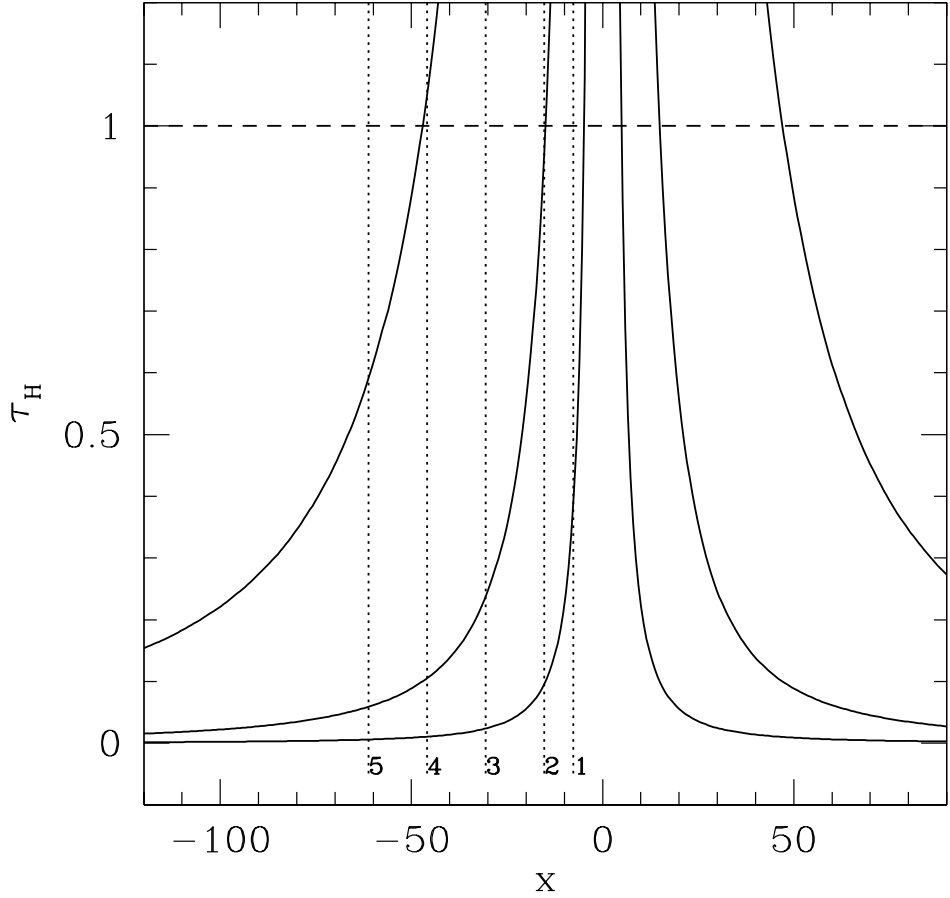,height=14cm}}
\caption
{Ly$\alpha$ scattering opacity for the cases with the line-centre
optical depths of $\tau_0= 10^5, 10^6, 10^7$, which are drawn
by solid lines from the innermost one. The vertical dotted lines
represent the effective scattering frequencies, and the number 
refers to the order of back-scattering with atoms in the supershell.}
\end{center}
\end{figure*}
%%%%%%%%%%%%%%%%%%%%%

The three peaks at the red part are composed of back-scattered photons.
Since the supershell is receding with respect to the incident photon,
scatterings usually occur at wing frequencies if the opacity at the 
frequencies is sufficiently large. Here we define the effective 
scattering opacity at a wing frequency $x$ by $\tau_e(x)$.
At the first scattering of Ly$\alpha$ photon with the expanding supershell,
the effective optical depth should be measured at $x_0+x_{\rm exp}$, i.e
$\tau_e(x_0+x_{\rm exp})$. Here we consider an input 
frequency $x_0 =0$, for simplicity. Note that $x_{\rm exp} < 0$.
If $\tau_e(x_0+x_{\rm exp}) \gg 1$, a back-scattering occurs 
at the wing frequency. Then, the back-scattered photons are redshifted 
to the frequency $x=x_0 + x_{\rm exp}$, and they will scatter 
in the other part of the supershell along the back-scattered direction.
If the supershell in that direction is optically thin,
then those back-scattered photons can escape freely to form a prominent 
peak. In fact, that part of the supershell is also receding with respect 
to the back-scattered photons, and so the effective opacity
now becomes $\tau_e(x_0+2x_{\rm exp})$. The opacity at this frequency 
is even smaller than that of previous scattering,
and the escape probability increases.
Hence some photons escape the system and form the first peak
at $x = x_0 + x_{\rm exp}$.
The photons reflected into slant directions 
are redshifted by an amount smaller than those to the normal
direction, and the optical depths in those direction becomes larger.
Hence the escape probability of those photons become smaller
than that of photons in the normal direction. 
These photons contribute to the broadening of each peak.
It is also noticeable in the figure that the primary peak is
not well matched with $x_{\rm exp}$, because its blue part is 
eroded by the scattering of very thick supershell whose
scattering centre is located at $x = -x_{\rm exp}$ in the figure.

Ly$\alpha$ photons repeat the above-mentioned back-scatterings 
until $\tau_e$ becomes small enough to permit
most of the photons to escape the system.
We see that photon's effective scattering frequency at the $N$-th
back-scattering is given by $2(N-1) x_{\rm exp}$, where $N > 1$.
The frequency of photons escaping associated with $N$-th back-scattering
is given by $(2N-3) x_{\rm exp}$.
When $N=1$, the situation is special, because the source is at rest.
In this case, each frequency is $x_{\rm exp}$ and $0$, respectively.

We illustrate in Figure~3 how the effective opacity 
of successive back-scatterings varies, where we consider
photons scattering back and forth along a diameter for simplicity.
The effective frequencies of Ly$\alpha$ photons successively
change, as is drwan by the vertical dotted lines in the figure.
The numbers in the box for each dotted line refers to
the order of back-scatterings denoted by $N$.
Let us consider the case for $\tau_0=10^6$ which corresponds
to the case in Figure~2. We can see that the effective opacity 
for $N=1$ is very large, and so most of Ly$\alpha$ photons 
are back-scattered. This means that 
there is a little amount of transmitted flux with $x=0$.
The second or $n=2$ back-scattering happens at wing frequencies 
much further away from the $N=1$ frequency, and so the escape 
probability is enhanced, but it is still not large enough
for most photons to escape the system. However, the emission 
peak forms at $x=x_{\rm exp}$, because the flux of incident Ly$\alpha$ 
photons is very large. In the next turn with $N=3$, the second peak
at $x=3x_{\rm exp}$ forms due to the process similar to the $N=2$ case.
When $N=4$, we expect that the other peak, but it appears very weak 
because most Ly$\alpha$ photons have already escaped from 
the system. From then on the effective opacities for the following 
back-scatterings are sufficiently small, but photons have been 
already exhausted. So no peaks at the higher frequencies appear 
in the emergent spectrum, as shown in Figure~2. 
We can summarize mathematically these explanation 
as follows.

\begin{equation}
I_1 (0) = I_0 ~ f_{\rm T} (x_{\rm exp}) \simeq 0,
\end{equation}
\begin{eqnarray}
I_2 (x_{\rm exp}) &=&
 I_0 ~ \left [ 1 - f_{\rm T} (x_{\rm exp})\right] f_{\rm T} (2x_{\rm exp}) \\
 &=& I_0 f_{\rm R} (x_{\rm exp}) f_{\rm T} (2x_{\rm exp}),
\nonumber
\end{eqnarray}
\begin{equation}
I_3 (3x_{\rm exp}) = I_0 ~ f_{\rm R} (x_{\rm exp}) f_{\rm R} (2x_{\rm exp}) f_{\rm T} (4x_{\rm exp}),
\end{equation}
\begin{equation}
I_4 (5x_{\rm exp}) = I_0 ~ f_{\rm R} (x_{\rm exp}) f_{\rm R} (2x_{\rm exp}) f_{\rm R} (4x_{\rm exp}) f_{\rm T} (6x_{\rm exp}),
\end{equation}
and so on. Generally
\begin{eqnarray}
&& I_N \left([2N-3]x_{\rm exp}\right) = I_0 f_{\rm R} (x_{\rm exp}) \\
  && ~~~~ \prod^{N-1}_{n=2} f_{\rm R}\left([2n-2]x_{\rm exp}\right) 
\cdot f_{\rm T} \left([2N-2]x_{\rm exp}\right),
\nonumber
\end{eqnarray}
where the transmitted fraction ($f_{\rm T}$) and the reflected 
fraction ($f_{\rm R}$) are approximately described in Eq.(2.32) and 
Eq.(2.33), respectively of Neufeld (1990). Here $I_0$ is the input flux,
and $I_N$ is the flux associated with the $N$-th back-scattering.

%%%%%%%% FIG 4
\begin{figure*}
\begin{center}
\parbox{1cm}{
\psfig{file=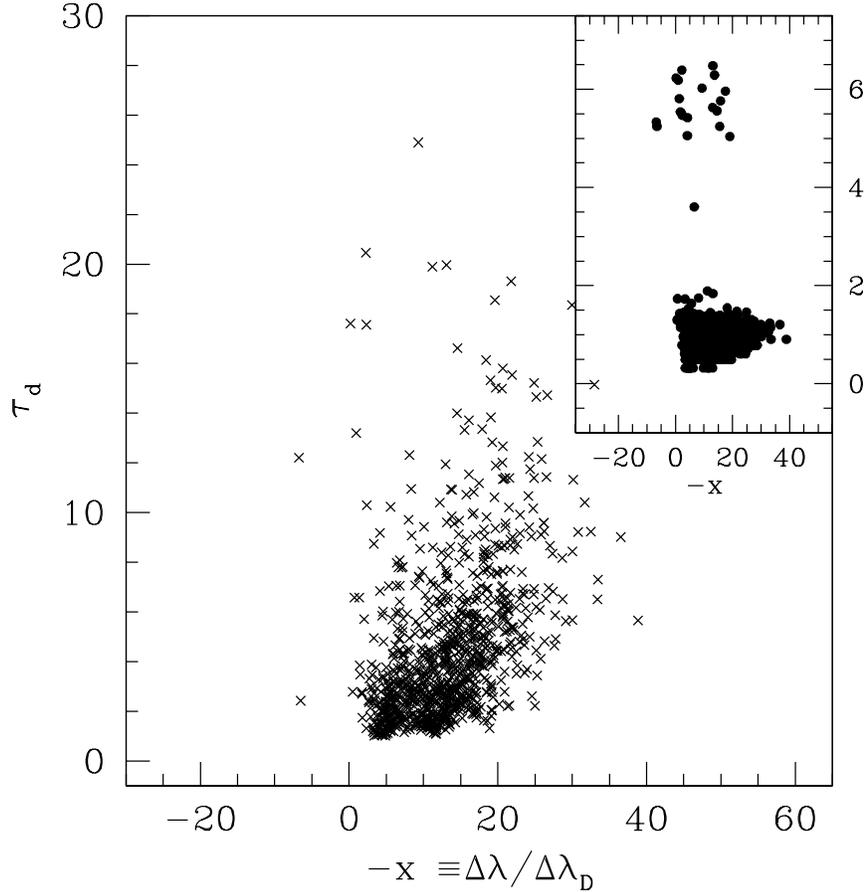,height=14cm}}
\caption
{Spectral distribution of the number of scatterings ($N_{\rm scat}$)
denoted by solid dots
and the total path lengths ($\tau_{\rm d}$) represented by crosses.
The calculation is performed for a model with the line-centre optical 
depth $\tau_0= 8.3\times 10^6$, 
the expansion velocity in units of thermal width $x_{\rm exp}=-7.65$,
and the FWHM of input profile $\Delta x=2$.}
\end{center}
\end{figure*}
%%%%%%%%%%%%%%%%%%%%%

In order to see how much emergent Ly$\alpha$ photons are affected
by dust extinction, we calculate the total path length of the
emergent photons through the medium. 
Here we adopt $N_{\rm HI} \simeq 2 \times 10^{20} \rm cm^{-2}$,
$b=40 \kms$, and $V_{\rm exp} \simeq 300 \kms$.
We also assume that dust is uniformly distributed in the supershell.

In Figure 4 we show the spectral distribution of both the total path length 
of Ly$\alpha$ photons in the supershell ($\tau_{\rm d}$) and 
the total number of scatterings ($N_{\rm scat}$).
The latter represented by solid circles in the inset is 
depicted in logarithmic scale, and shows clearly two groups. 
One group is located at the upper domain with $\log_{10} N_{\rm scat} \sim 6$,
and they are photons having experienced excursions.
The rest in the lower part with $\log_{10} N_{\rm scat} \approx 1$
are those having escaped by a number of back-scatterings.

The crosses represent, in units of $\tau_0$, the total path length
along which a Ly$\alpha$ photon traversed in the supershell. 
We can see in the figure that most of emergent photons
escape by back-scattering, which can be inferred from
the small number of scatterings. It is also noticeable that
the total path lengths are very diverse in the spectral distribution
and most of photons traverses rather short path.
Since there is no frequency-dependence in path lengths except for 
a weak trend of monotonous increases, the redder peaks 
in emergent profiles from dusty media can 
be destroyed at most a little bit more than the bluer peaks.
This is puzzling because all the observed Ly$\alpha$ profiles
shows single peak which must be the $N=2$ peak in our calculations.
The solution of this puzzle may lie in the spatial distribution
of dust in the system. If the inner radius of the dust shell is
is smaller than that of neutral hydrogen, the dust extinction
of the higher order peak can increase. 
We are now undertaking this subject,
and the results will be published in the forthcoming paper.

\section{Dependence on the Physical Parameters}

The emergent profiles depends on a few parameters of scattering
medium in our model. First of all, the turbulence is out of
our interest, because they are constrained by observations
and the emergent profile is dominantly formed by kinematics
of the supershell in our case. Hence here we investigate
how the Ly$\alpha$ peaks depend on the optical depths and 
the expansion velocity of the supershell.

\subsection{Effects of supershell's opacity}

%%%%%%%% FIG 5
\begin{figure*}
\begin{center}
\parbox{1cm}{
\psfig{file=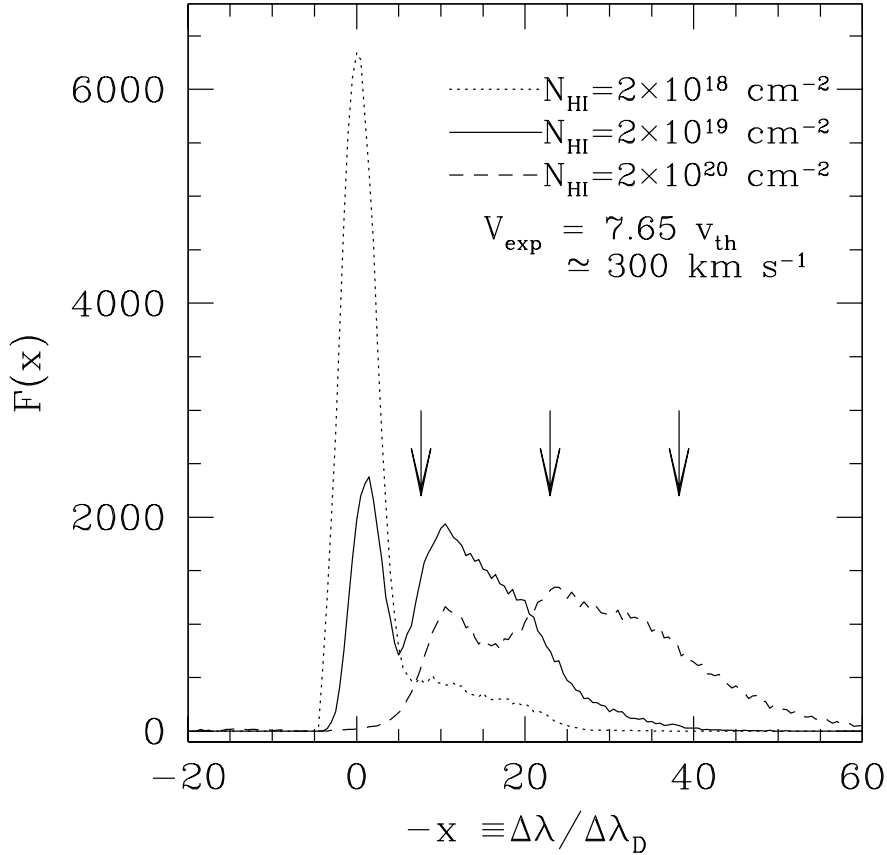,height=14cm}}
\caption
{Emergent Ly$\alpha$ profiles for three cases in which the supershells
have different line-centre optical depths but their expansion velocities
are fixed. The parameters are given in the box, and the arrows refers
to the location of possible emission peaks formed by back-scatterings.}
\end{center}
\end{figure*}
%%%%%%%%%%%%%%%%%%%%%

We present in Figure~5 the emergent profiles for a few cases with different
line-centre optical depths, where the expansion velocity of the supershell
is fixed to be $x_{\rm exp}=-7.65$. The FWHM of input profiles
is also set to be $\Delta x = 2$, and the Doppler parameter of
the supershell is set to be $b=40\kms$. The supershell is assumed to
have no dust, and the number of input photons are set to be equal
for each calculation or $N_{\rm ph}=80,000$.
The arrows in the figure represent the peaks corresponding to
$x_{\rm exp}$, $3x_{\rm exp}$, and $5x_{\rm exp}$.
When the opacity of the supershell is not sufficiently large
($N_{\rm HI}=2\times 10^{18},~2\times 10^{19}$),
the input photons are transmitted without scatterings
and the peak at $x \simeq 0$ forms, whose FWHM
increases by a small amount compared to that of input profile.
We can see in the figure that the back-scattered photons 
form the primary peak at $x\simeq x_{\rm exp}$,
the secondary peak at $x\simeq 3x_{\rm exp}$, and
the tertiary peak at $x\simeq 5x_{\rm exp}$.
As the opacity of the supershell decreases, the peaks at higher
frequencies disappear accordingly.
We have explained the mechanism of those variations in the previous
section.

\subsection{Effects of Kinematics}

%%%%%%%% FIG 6
\begin{figure*}
\begin{center}
\parbox{1cm}{
\psfig{file=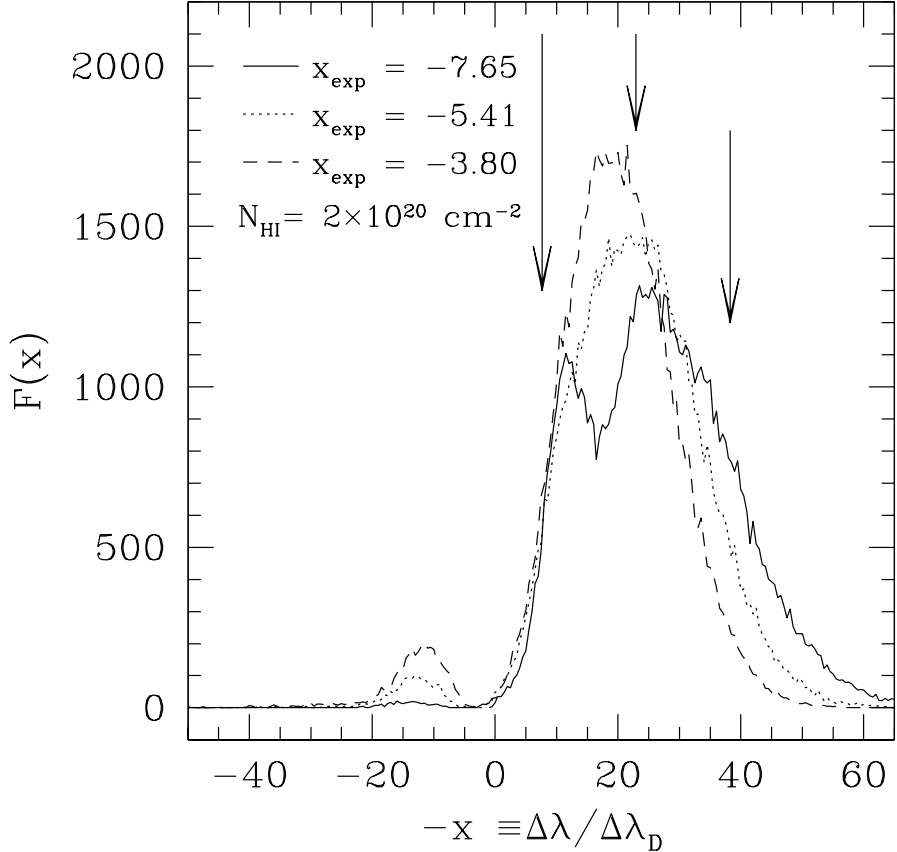,height=14cm}}
\caption
{Emergent profiles of Ly$\alpha$ for several cases with different 
expansion velocities.
The horizontal axis refers to the wavelength in units of thermal width,
and the vertical axis represents the number flux of emergent profiles.
We assume that the input profiles have the same 
Gaussian input profile with $\Delta x=2$ FWHM and the same number flux 
of photons or 80,000. The column density of neutral hydrogen 
in the supershell is also fixed or $N_{\rm HI}=2\times10^{20} \cm^{-2}$.
The meaning of each line is shown in the box, and 
the arrows represent the frequencies
for the primary, secondary, and tertiary peaks for $x_{\rm exp}=-7.65$.}
\end{center}
\end{figure*}
%%%%%%%%%%%%%%%%%%%%%

We also investigate the effects of the expansion velocity of 
the supershell. The results of Monte Carlo calculations are
shown in Figure~6. We can see that the emission peaks 
gradually merge into a single peak as $|x_{\rm exp}|$ decreases.
This is because the supershell behaves like a spherical 
medium which is static with respect to the Ly$\alpha$ source
as $|x_{\rm exp}| \rightarrow 0$.
In an extremum with $|x_{\rm exp}| < 1$, the emergent profiles
will have symmetric double peaks which is same to those 
given in Ahn et al. (2002). If $|x_{\rm exp}|$ is much larger 
than unit, then we can see the emission peaks
of back-scattered photons in the emergent profile.
However, as $|x_{\rm exp}| \rightarrow 1$,
the frequency gaps between those peaks become close, 
and they eventually mixed up to be a single broad peak.
We can also expect a heavy extinction for small $|x_{\rm exp}|$ cases.

It is also noticeable that the similar problem for the contraction
supershell has a exact mirror-symmetry to the expanding supershell. 
Hence it is evident that we obtain the profiles 
to the origin $x=0$ for the contracting case.

\section{Summary}

We have investigated the formation of P-Cygni type Ly$\alpha$
by a supershell surrounding a star-forming region and outflowing 
in a bulk manner with expansion velocities of $V_{\rm exp} \sim 100-300 \kms$.
We have modified the Monte Carlo code developed in the previous paper
(Ahn et al. 2001), and studied
in detail the Ly$\alpha$ line transfer in the expanding supershell
consists of neutral hydrogen gas. The Ly$\alpha$ line-centre
optical depth of the supershell is very large and of 
order $\tau_0 \sim 10^5-10^7$.
Its turbulence motion is assumed to be about $40\kms$ 
in accordance with observations.

We found that the back-scatterings play a crucial role in the 
formation of asymmetric Ly$\alpha$ profiles. Most of Ly$\alpha$ 
photons escape the system efficiently through back-scatterings.
Whenever it is scattered back by the expanding supershell, 
photon's frequency is redshifted successively.
This process promotes the redshift of Ly$\alpha$ photons.
It is also naturally expected that the collapsing supershell 
promotes blueshift of Ly$\alpha$ photons,
and the emergent profile for that case is simply 
mirror-symmetric to that of the expanding case.
The emission peak formed by the $N$-th back-scattering
with the supershell has a frequency of $(2N-3)x_{\rm exp}$,
where $N>1$. The number of peaks and their relative flux ratios
in emergent profiles are determined by the interplay 
between the line-centre optical depths and 
the expansion velocities of supershells.

Recently optical nebular 
lines of about a dozen of Lyman break galaxies 
have been obtained by using near-IR spectrographs and large 
telescopes (Pettini et al. 2001; Teplitz et al. 2000a; Teplitz et al. 2000b).
The Ly$\alpha$ spectra of a large number of primeval galaxies 
can be found in the literature (Steidel et al. 1996; Steidel et al. 1999;
Lowenthal et al. 1997; Frye et al. 2002),
which are reviewed in Stern \& Spinrad (1999).
Their UV spectra and Ly$\alpha$ emission lines show similar characteristics 
to those of nearby starburst galaxies.
Their P-Cygni type Ly$\alpha$
emission lines are always accompanied by nebular lines
blueshifted with respect to Ly$\alpha$.
We have showed that these asymmetric Ly$\alpha$ can be formed by
back-scatterings of Ly$\alpha$ photons by an expanding supershell.

The quantitative measurement of star-formation rates of starburst 
galaxies is one of the most crucial issues in modern cosmography.
It is suggested that the expansion of surrounding media can help 
Ly$\alpha$ photons escape the star-forming regions more freely.
In light of our works, it is very interesting to see that
nearly all the observed P-Cygni type Ly$\alpha$ emissions from 
starburst galaxies do not show any higher order peaks
which are formed by back-scatterings.
This can certainly be attributed to dust extinction.
However, it is very subtle
problem if we want to quantify the escape fraction of Ly$\alpha$
photons for the case of the dusty and expanding supershell.
Hence, we believe that very careful consideration of both abundance 
and spatial distribution of dust is needed to establish 
a realistic model describing the star-forming activity 
in starburst galaxies whether it is nearby or primeval.
We are now undertaking this topic, and its results will be
reported in a forthcoming paper.

\section*{Acknowledgments}
This work was done as a part of doctoral dissertation of SHA
at the School of Earth and Environmental Sciences of Seoul National University
financially supported by Brain Korea 21 of the Korean Ministry of Education.
Also this work is completed
with financial support of Korea Institute for Advanced
Study, which is gratefully acknowledged. HML was supported by KOSEF Grant
No. R01-1998-00023.

\end{document}